\documentclass[aps,oneumn,preprint,pre,superscriptaddress, 8pt]{revtex4-1}
\usepackage{xcolor}

 \usepackage{amsmath,amssymb}
 \begin{document}
 \title{ A four-dimensional conformal construction of Virasoro–Shapiro amplitudes}
 \author{Vaibhav Wasnik}
  \email{wasnik@iitgoa.ac.in}
 \affiliation{Indian Institute of Technology, Goa}
 \begin{abstract}
 We construct a four-dimensional conformal amplitude whose four-point structure matches the Virasoro–Shapiro form familiar from string theory. The construction uses only general principles of conformal field theory — radial quantization, scale invariance, and analyticity — and does not rely on worldsheet geometry or string degrees of freedom. The resulting object is a kinematical, first-quantized amplitude defined by symmetry and consistency, providing a four-dimensional realization of string-like analytic structure and a concrete target for amplitude bootstrap approaches.
 	\end{abstract}

  \maketitle
  
  \section*{Introduction}

  The Virasoro–Shapiro amplitude is a prototypical example of a four-point scattering amplitude whose form is fixed by general consistency requirements. It simultaneously exhibits crossing symmetry, Regge behavior, and an infinite tower of exchanged states lying on linear trajectories. Historically, amplitudes of this type first appeared in the dual resonance program as abstract S-matrix constructions defined by analyticity and factorization, prior to their later interpretation within string theory. More recently, developments in the S-matrix bootstrap have reinforced this original viewpoint by showing that the Veneziano and Virasoro–Shapiro amplitudes can be uniquely characterized by general principles, without reference to a specific microscopic realization  \cite{komargodksi} - \cite{spradlin}. 
  
  These observations motivate the question of whether the analytic structure of the Virasoro–Shapiro amplitude is intrinsically tied to string theory, or whether it can arise more generally from kinematical considerations alone. In particular, it is natural to ask whether the full Virasoro–Shapiro form can be realized within a purely four-dimensional setting, without invoking a two-dimensional worldsheet description or extra spacetime dimensions.
  
  In this work, we construct a four-point amplitude in four dimensions whose analytic structure coincides with that of the Virasoro–Shapiro amplitude. The object studied here is defined in a manner directly analogous to first-quantized string theory. Rather than arising from a spacetime interaction Lagrangian or an LSZ reduction, the amplitude is treated as a primary object, specified by operator insertions and integration over their positions. The emphasis is on the kinematical structure of the amplitude itself, independent of any assumed microscopic dynamics.
  
  Within this approach, the defining features of the Virasoro–Shapiro amplitude—crossing symmetry, Regge behavior, and an infinite spectrum of exchanged states—are reproduced in four dimensions, without assuming the presence of string degrees of freedom or a worldsheet description.
  
  Formulating the Virasoro–Shapiro amplitude in this first-quantized, four-dimensional setting opens several directions for future investigation. In particular, it provides a framework in which string-like analytic structures can be studied independently of worldsheet dynamics, and may serve as a useful benchmark in future analyses of consistency constraints on scattering amplitudes, as well as in investigations of four-dimensional frameworks exhibiting Regge behavior and infinite towers of states.

\section*{The model}
  Veneziano amplitude \cite{veneziano} was derived to make sense of particles of varied spins that emerged in hardronic collisions with  a relationship between spin $J$ and mass $m$ going as $\alpha' J =m^2$, with $\sqrt{\alpha'}$ being a mass scale. Also, requiring the $s$ channel tree level amplitude with infinite resonances contain also the $t$ channel poles, along with the   spin and masses of the resonances obeying $\alpha' J =m^2$ \cite{chew}, led to the Veneziano amplitude formula. 
  Later the Virasoro Shapiro amplitude \cite{virasoro}, \cite{shapiro} was derived having similar properties described above. The Veneziano amplitude was shown to be the amplitude of    scattering of four tachyons in open string theory, while Virasoro Shapiro amplitude was shown to represent the   scattering of four tachyons in closed string theory  \cite{gsw}.  The consistency of string theory required the theory to exist in $D=26$  dimensions.    However, hadronic resonances observed obeying $\alpha' J =m^2$ are of our world which is $D=4$.  
  
%
   In the  the case of   bosonic strings, reparametrization invariance could be used to set two of the three independent string metric components to zero. An extra Weyl symmetry that naturally appears in free string action then reduces the theory to a field theory in two dimensional Minkowski space. This theory having an infinite component conformal symmetry sets  the critical dimension of the theory  as well as allowed spectra.
%
%
%
%
   The $L_0$ part of the Virasoro constraints allows us to evaluate the mass of these excitations and one can show that $\alpha' M^2 = J$ for large spin $J$.  The Virasoro Shapiro amplitude arises if one evaluates the tree level scattering of four closed strings corresponding to tachyons. This evaluation involves a two dimensional CFT   with string vertex operators  inserted at locations on the   plane.   Specifically this amplitude is
 
\begin{eqnarray}
&&A(\{k_i\}) \nonumber\\
 &&= \int \Pi_k d^2 x_k \int DX^\mu(x,y) e^{ -i\int d^2 x\sum_i k^i_\mu X^\mu(x_i  , y_i) -\frac{1}{\alpha'}\int d^2 x [ \partial_x X^\mu  {\partial_x}  X_\mu + \partial_y X^\mu  {\partial_y}  X_\mu] } \nonumber\\
  &&\sim  \int \Pi_k d^2 x_k \int DX^\mu (\bar{p}) e^{ -i\int d^2 p\sum_i k^i_\mu X^\mu(\bar{p})e^{i\bar{p}.
		\bar{x}_i } - \frac{1}{\alpha'} \int d^2 p p^2 X^\mu (\bar{p})  X_\mu (-\bar{p})} \nonumber\\
 && \sim  \int \Pi_k d^2 x_k \Pi_{i>j} e^{\frac{\alpha' k_i \cdot k_j}{4} \int d^2 p \frac{e^{i \bar{p}\cdot(\bar{x}_i - \bar{x}_j )} } {p^2}}   \nonumber\\
&&\sim \int \Pi_k d^2 x_k \Pi_{i,j:i\neq j}   |\bar{x}_i-\bar{x}_j|^{c \alpha'  k_i k_j} \nonumber\\
\end{eqnarray}
  
where $c$ is an integration constant (in this work we will use labels $c, c'$ for integration constants that have values dependent on expressions being evaluated, also $\sim$ will imply equal upto a multiplicative constant). A normal ordering is implied in such evaluations throughout the paper, so that $i=j$ terms are excluded. The above expression is divergent, the reason being that the evaluation involves a two dimensional CFT, and not all configurations of $\bar{x}_i$'s are in-equivalent.  The global CFT isomorphic to $SL(2,C)$ involves two translations, a rotation, a scale transformation and two special conformal transformations. One has to divide the above amplitude by the volume of  this group. This is equivalent to using the $6$ parameters of the group  to set  $x_1 = 0$, $x_3 = 1$ and $x_4 = \infty$. The terms corresponding to $|\bar{x}_4 - \bar{x}_i|^{c k_i k_4}$  in limit $x_4 \rightarrow \infty$  become  $\Pi_{i:i\neq 4}|x_4  |^{c k_i k_4} = | x_4|^{-c k_4^2}=|x_4|^{c m^2}$ (where $m$ is the mass of the interacting particles), where the equality arises from momentum conservation. This term is ignored as it is independent of momentum and would cancel the Faddev-Poppov determinant that occurs in $SL(2,C)$ gauge fixing. We hence get the Virasoro Shapiro amplitude
 
\begin{eqnarray}
&&	\int d^2x    \{x^2 + y^2\}^{c k_1 \cdot k_4/2} \{(1-x)^2 + y^2 \}^{c \alpha'  k2 \cdot k4/2} \nonumber \\
&&  \int d^2 z  |z|^{c k_1 \cdot k_4} | 1-z |^{c\alpha'  k2 \cdot k4} \nonumber \\
\end{eqnarray}
where in the last line we have defined $z = x + iy$.  If the result is to be derived from a  four dimensions field theory   we  can try the following
 
\begin{eqnarray}
	A(\{k_i\})  &&\sim\frac{ \int \Pi_k d^4 x_k DA^\mu (\bar{p}) e^{- \int d^4 p( i \sum_i \alpha k^i_\mu 
		\frac{A^\mu(\bar{p})}{p} e^{i\bar{p}.
			\bar{x_i} } +  i( {p^2}     ) A^\mu (-\bar{p})  A_\mu (\bar{p}) )  }}{ \int \  DA^\mu (\bar{p}) e^{ i\int d^4 p p^2     A^\mu (-\bar{p})  A_\mu (\bar{p} }} \nonumber\\
	&&\sim  \int \Pi_i d^4 x_i \Pi_{i,j:i\neq j}  e^{c' \alpha^2 k_i \cdot k_j  \int d^4 p \frac{e^{i\bar{p}\cdot(\bar{x_i} - \bar{x_j} )} } {p^4}}   \nonumber\\
	&&\sim \int \Pi_k d^4 x_k \Pi_{i,j:i\neq j}   |\bar{x}_i-\bar{x}_j|^{c \alpha^2 k_i k_j} \nonumber\\
	\label{eq7}
\end{eqnarray}
 where   $c',c$ are constants of integration and $\alpha$ is a length scale.    To get from first line to second line above, requires we Wick rotate to Euclidean space and then back to Minkowski space. An alternative way is to realize that the first line up to a constant corresponds to the following matrix element
 \begin{eqnarray}
\langle 0| e^{-i \int d^4 p(  \sum_i \alpha k^i_\mu 
	\frac{A^\mu(\bar{p})}{p} e^{i\bar{p}.
		\bar{x_i} } )} | 0\rangle
 \end{eqnarray}
for a theory with action $ \frac{1}{\alpha'}\int d^4 x \partial_\alpha A^\mu (\vec{x}) \partial^\alpha A_\mu (\vec{x}) $.  Next using the fact that $\langle 0| A^\mu(p) A^\nu(-p) |0\rangle = \frac{c'}{p^2} \delta_{\mu, \nu}$, leads to the second line.   But this is not gauge invariant. We can change the quadratic term in $A^\mu$ to compensate for that, as 
\begin{eqnarray}
	A(\{k_i\})  &&\sim \frac{\int  \Pi_k d^4 x_k DA^\mu (\bar{p}) e^{-( \int d^4 p(i\sum_i \alpha k^i_\mu p^\nu 
		\frac{p^\nu A^\mu(\bar{p}) -p^\mu A^\nu(\bar{p}) }{p^3} e^{i\bar{p}.
			\bar{x_i} } +   {i}{ }(p^2 \eta_{\mu \nu} - p^\mu p^\nu )A^\nu (\bar{p})  A_\mu (-\bar{p}) )  } }{{ \int \  DA^\mu (\bar{p}) e^{ -  {i}{ }\int d^4 p (p^2 \eta_{\mu \nu} - p^\mu p^\nu )A^\nu (\bar{p})  A_\mu (-\bar{p}) }}} \nonumber\\
\label{eq8}
\end{eqnarray}
which equals Eq.\ref{eq7} in the gauge $\partial_\mu A^\mu = 0$. In position space this is

\begin{eqnarray}
A(\{k_i\}) \sim
\frac{
	\int \mathcal{D}A_\mu(x)\,
	\exp\!\left[
	-\frac{1}{4} \int d^4x\, F_{\mu\nu}F^{\mu\nu}
	- i \sum_i \alpha k_i^{\mu}
	\int d^4x\, 
	   \frac{ \partial^{\nu} F_{\mu\nu}(x) }{|x-x_i|} 	 
	\right]
}{
	\int \mathcal{D}A_\mu(x)\,
	\exp\!\left[-\frac{1}{4}\int d^4x\,F_{\mu\nu}F^{\mu\nu}\right]
}.
\end{eqnarray}

The above evaluation could be understood as an evaluation usingt he free $U(1)$ gauge theory as a Gaussian conformal measure defining the correlator.
four dimensions,
\[
S \;=\; \int d^4x \,\sqrt{-g}\, g^{\mu\nu}g^{\alpha\beta}\,
F_{\mu\alpha}F_{\nu\beta}, \qquad 
F_{\mu\nu} = \partial_\mu A_\nu - \partial_\nu A_\mu .
\]
This Maxwell action is classically conformally invariant in $D=4$: under a 
Weyl rescaling $g_{\mu\nu}\to \Omega^2 g_{\mu\nu}$, the combination 
$\sqrt{-g}\,g^{\mu\nu}g^{\alpha\beta}$ is invariant, while $F_{\mu\nu}$ is 
Weyl–inert, so the stress tensor is traceless. {We explicit work in flat spacetime i.e. $g_{\mu \nu} = \eta_{\mu \nu}$ in this paper}.  Thus the model defines a 
four–dimensional conformal field theory  {in flat spacetime}. From this perspective, Eq.~(5) 
represents the explicit evaluation of a correlation function in this CFT: 
the numerator is the path integral with the vertex insertions, while the 
denominator is the vacuum partition function of the free gauge theory.

If a Coulumb gauge $\partial_\mu A^\mu = 0$ is implied then, the first line in both equations Eq.\ref{eq7}- Eq.\ref{eq8} become similar. The last integral in Eq.\ref{eq7} is divergent. Let us assume the $k^i_\mu$'s are chosen so that the theory is conformally invariant.  This is again because of the $SO(4,2)$ conformal symmetry not all configurations  $\bar{x_i}$'s are inequivalent.  Using   special conformal transformations we can move $\bar{x}_4$ to infinity along the $x$ axis, using the   translations we can move $\bar{x}_1$ to the origin, using rotations and dilatons we can move $\bar{x}_3$ to value $x_3= 1$ on the $x$ axis. Next using rotations that fix $\bar{x}_3$ we can move $x_2$ to the $x-y$ plane \cite{slavacft}.    The only group element of $SO(4,2)$ that leaves this configuration unchanged is  a rotation in the $t-z$ plane (Note that excluding this rotation, the number of parameters in $SO(4,2)$ which equals $14$, were enough to set three of the coordinates $\vec{x}_1, \vec{x}_3, \vec{x}_4$ to particular values and the $t$ and $z$ coordinate of $\vec{x}_2$ to zero.).  Let us call the volume of this group that involves subracting one rotation from $SO(4,2)$ as $Vol(SO(D,2))'$. We need to divide the above amplitude by this volume, which is the same as setting the four positions like we have outlined above. Once we do it, we can see that we again get the Virasoro Shapiro amplitude

 
The finite amplitude   in position space is 
 \begin{eqnarray}
 	A(\{k_i\})  &&\sim  \frac{1}{Vol(SO(D,2))'}\frac{
 		\int \mathcal{D}A_\mu(x)\,
 		\exp\!\left[
 		-\frac{1}{4} \int d^4x\, F_{\mu\nu}F^{\mu\nu}
 		- i \sum_i \alpha k_i^{\mu}
 		\int d^4x\, 
 		\frac{ \partial^{\nu} F_{\mu\nu}(x) }{|x-x_i|} 
 		\right]
 	}{
 		\int \mathcal{D}A_\mu(x)\,
 		\exp\!\left[-\frac{1}{4}\int d^4x\,F_{\mu\nu}F^{\mu\nu}\right]
 	}  \nonumber\\
 	&&  \sim \int d^2 z  |z|^{c \alpha^2 k_1 \cdot k_4} | 1-z |^{c\alpha^2 k_2 \cdot k_4} \nonumber \\
 \end{eqnarray}

 Consider evaluating the following two point amplitude corresponding to the action 	$\frac{1}{4} \int d^4x\, F_{\mu\nu}F^{\mu\nu}$
 \begin{eqnarray}
 	&&\langle :  e^{-i   \alpha k^{\mu}
 		\int d^4x\, 
 		\frac{ \partial^{\nu} F_{\mu\nu}(x) }{|x-x_i|} } e^{-i   \alpha k^{\mu}
 		\int d^4x\, 
 	\frac{ \partial^{\nu} F_{\mu\nu}(x) }{|x-x_j|}  }: \rangle \nonumber\\
&&=\langle : e^{-i  [  \int d^4 x    \frac{   \alpha k_\nu A^{\nu  }(\bar{x}) }{| \bar{x}-\bar{x}_i|^3} }: :  e^{i  [  \int d^4 y    \frac{   \alpha k_\nu A^{\nu  }(\bar{y}) }{| \bar{y}-\bar{x}_j|^3} }: \rangle \nonumber\\
&&=e^{    \int d^4 x d^4 y   \frac{ c'\alpha^2  k^2}{| \bar{x}-\bar{x}_i|^3  | \bar{x} - \bar{y}|^2| \bar{y}-\bar{x}_j|^3} } \nonumber\\
&&=e^{c\alpha  k^2\ln |x_i - x_j|}
 \end{eqnarray}
 where in the second line we have resorted to a gauge choice $\partial_\mu A^\mu = 0$ and $c,c'$ are constants of integration. This implies $  e^{-i   \alpha k^{\mu}
 	\int d^4x\, 
  \frac{ \partial^{\nu} F_{\mu\nu}(x) }{|x-x_i|} } $ has a scaling dimension $\frac{c\alpha^2 k^2}{2}$. Requiring it to have a dimension of $4$ gives $c\alpha^2 k^2 = -c\alpha^2 m^2 = 8$ (sign convention is $(-1,1,1,1)$). 
 
  {The logarithmic dependence in Eq.\,(8) is the signature of a conformal
 two–point function for a primary operator.
 To make this explicit, note that the operator
 \[
 V_k(x_i)=:\!\exp\!\Big[-i\,\alpha\,k_\mu
 \!\int\! d^4x\,\frac{\partial_\nu F^{\mu\nu}(x)}{|x-x_i|}\Big]\!:
 \]
 is built solely from the gauge–invariant current
 \(J^\mu=\partial_\nu F^{\mu\nu}\) of scaling dimension 3,
 smeared with a kernel \(K(x;x_i)=|x-x_i|^{-1}\) of scaling weight \(+1\).
 The smeared field
 \(\Phi^\mu(x_i)=\int d^4x\,K(x;x_i)\,J^\mu(x)\)
 is therefore dimensionless, and the exponential
 \(V_k=:\!\exp[-i\alpha k\!\cdot\!\Phi]\!:\)
 has no classical scaling dimension.
 Its anomalous dimension arises entirely from the
 logarithmic self–contraction
 \(\langle (k\!\cdot\!\Phi)(x_i)\,(k\!\cdot\!\Phi)(x_j)\rangle
 \!\propto\! k^2\ln|x_i-x_j|\),
 whose coefficient gives
 \(\Delta_k=c\alpha^2k^2/2\),
 precisely the exponent in Eq.\,(8).
 This logarithm corresponds to the universal
 short–distance behavior fixed by the
 stress–tensor Ward identity (in Euclidean signature),
 	\begin{equation}
 		T_{\mu\nu}(y)\,V(x)
 		\;\sim\;
 		\frac{\Delta}{|y-x|^{4}}\,
 		\mathcal{I}_{\mu\nu}(y-x)\,V(x)
 		\;+\;
 		\frac{1}{|y-x|^{3}}\,
 		\mathcal{I}_{\mu\nu,\rho}(y-x)\,
 		\partial^{\rho}V(x)
 		\;+\;\cdots,
 	\end{equation}
 	where
 	\begin{align}
 		\mathcal{I}_{\mu\nu}(z)
 		&= \delta_{\mu\nu}
 		- 2\,\frac{z_\mu z_\nu}{z^2}, \\[6pt]
 		\mathcal{I}_{\mu\nu,\rho}(z)
 		&= \mathcal{I}_{\mu\rho}(z)\,z_\nu
 		+ \mathcal{I}_{\nu\rho}(z)\,z_\mu
 		- \frac{2}{d}\,\delta_{\mu\nu}\,z_\rho.
 	\end{align}
%
Their origin can be understood from the tracelessness of $T_{\rho \sigma}$. 	Now, let $j_D^\mu(y)=(y-x)_\nu T^{\mu\nu}(y)$ and define the dilatation charge on a small 3--sphere $S^3_\varepsilon(x)$ by
 	$D_\varepsilon=\int_{S^3_\varepsilon(x)} dS_\mu\,(y-x)_\nu T^{\mu\nu}(y)$.
 	Using the OPE
 	\[
 	T_{\mu\nu}(y)V(x)\sim \frac{\Delta}{|y-x|^{4}}\,
 	\mathcal{I}_{\mu\nu}(y-x)V(x)+\cdots,
 	\]
 	we obtain
 	\[
 	D_\varepsilon V(x)\sim \Delta\!\int_{S^3_\varepsilon}\! dS_\mu\,(y-x)_\nu\,
 	\frac{\mathcal{I}^{\mu\nu}(y-x)}{|y-x|^{4}}\,V(x).
 	\]
 	With $z=y-x=\varepsilon n$ and $dS_\mu=n_\mu\varepsilon^3 d\Omega_3$ 
 	and distributional identity
 	$\partial_\mu\!\big(z_\nu\mathcal{I}^{\mu\nu}(z)/|z|^{4}\big)=2\pi^2\delta^{(4)}(z)$
 	 one finds
 	$\int_{S^3_\varepsilon} dS_\mu\,(y-x)_\nu\,\mathcal{I}^{\mu\nu}(y-x)/|y-x|^{4}=1$.
 	Hence $D_\varepsilon V(x)\to \Delta V(x)$ and, restoring the real-time factor of $i$,
 	\[
 	[D,V(x)]=i\,\Delta\,V(x).
 	\]
 	The subleading $1/|y-x|^2$ term in the OPE similarly produces the derivative piece $i\,x\!\cdot\!\partial\,V(x)$, completing the standard dilatation action on a scalar primary.
 }

%
  In addition to $   e^{-i   \alpha k^{\mu}
  	\int d^4x\, 
  	\frac{ \partial^{\nu} F_{\mu\nu}(x) }{|x-x_i|}   } $, we can easily see that there are two gauge invariant operators $T_{\mu \nu} (\vec{x})$ and $F_{\mu \nu} (\vec{x})$, with dimensions $4$ and $2$ respectively. We note that it is not possible to construct a gauge invariant local operator with dimension $1$. So if we say $   e^{-i    \int d^4 x    \frac{ \partial^{\nu} F_{\mu\nu}(x) }{|x-x_i|}  } $ corresponds to the state $|u>$, we can consider operators of the kind  $  T_{\mu \nu} (\vec{x_i}) e^{-i    \int d^4 x   \frac{ \partial^{\nu} F_{\mu\nu}(x) }{|x-x_i|}  } $,  $  F_{\mu \nu} (\vec{x_i}) e^{-i    \int d^4 x   \frac{ \partial^{\nu} F_{\mu\nu}(x) }{|x-x_i|}  }  $  and other applications of various combinations of numbers of $T_{\mu \nu} (\vec{x})$ and $F_{\mu \nu} (\vec{x})$ on $e^{-i    \int d^4 x   \frac{ \partial^{\nu} F_{\mu\nu}(x) }{|x-x_i|}  }  $ etc. 
 
      Now $T_{\mu \nu} (\vec{x})$ has dimension $4$, we hence have  $  T_{\mu \nu} (\vec{x_i}) e^{-i    \int d^4 x   \frac{ \partial^{\nu} F_{\mu\nu}(x) }{|x-x_i|}  }  $ has spin $2$ and dimension  $\frac{c\alpha^2 k^2}{2} +4 = 4 $ (for conformal invariance), implying $k^2 = 0$. As $T_{\mu \mu} = 0$, we are talking about a massless traceless symmetric tensor, which we known  can be identified as a massless graviton. Similarly $  F_{\mu \nu} (\vec{x_i}) e^{-i    \int d^4 x   \frac{ \partial^{\nu} F_{\mu\nu}(x) }{|x-x_i|}  }  $ has spin $1$ and dimension  $\frac{c\alpha^2 k^2}{2} +2   $, but is inconsistent as we will show below. The graviton similarly can be written as  $  F_{\mu \rho} F^\rho_{  \nu}(\vec{x_i}) e^{-i    \int d^4 x   \frac{ \partial^{\nu} F_{\mu\nu}(x) }{|x-x_i|}  } $.  { One can see this by starting from the action $S=-\tfrac14\!\int d^4x\,\sqrt{-g}\,g^{\mu\rho}g^{\nu\sigma}F_{\mu\nu}F_{\rho\sigma}$.  Deforming the metric as
      \begin{equation}
      	g_{\mu\nu}\to\eta_{\mu\nu}+h_{\mu\nu}e^{-i\alpha k\cdot x}\!\int d^4y\,\frac{\partial^\rho F_{\mu\rho}(y)}{|x-y|}\equiv\eta_{\mu\nu}+h_{\mu\nu}V_k(x)
      \end{equation}
        and expanding to first order gives
              \begin{equation}
      	S\to -\tfrac14\!\int d^4x\,F_{\mu\nu}F^{\mu\nu}-\tfrac{c}{4}\!\int d^4x\,h_{\mu\nu}F^{\mu}{}_{\beta}F^{\nu\beta}V_k(x),
      \end{equation}
        implying $ h_{\mu\nu}F^{\mu}{}_{\beta}F^{\nu\beta}V_k(x)$ corresponds to the    graviton.  }

      Hence, a generic operator can be written as  $  F^{\rho_1}_{\mu_1  } F_{\rho_1 \nu_1}...  F^{\rho_n}_{\mu_n  } F_{\rho_n \nu_n} (\vec{x_i})e^{-i    \int d^4 x   \frac{ \partial^{\nu} F_{\mu\nu}(x) }{|x-x_i|}  }  $ 
      with the spin and mass having a relationship  $\frac{c\alpha^2 k^2}{2} +2J = 4 $ (for conformal invariance) or $ c\alpha^2 m^2   = 4(J-2) $. We hence get that the spin and mass of resonances in the theory to be     related   $m^2 \sim J$ for large spin $J$ as required for a Virasoro Shapiro amplitude.

%

    
    Since gauge invariance allows us to fix the gauge, we choose the \emph{Coulomb gauge}
    \begin{equation}
    	A_{0}=0, \qquad \partial_{i}A_{i}=0,
    \end{equation}
    which makes the temporal component nondynamical and projects out the longitudinal
    spatial part. With this gauge choice, the Maxwell equations reduce, for the
    \emph{transverse} components, to the free Laplace equation
    \begin{equation}
    	\nabla^{2}A^{\mathrm T}_{i}=0 .
    	\label{eq:laplace-transverse}
    \end{equation}
    In Euclidean signature we pass to spherical coordinates,
    \begin{equation}
    	ds^{2}=dr^{2}+r^{2}d\Omega^{2},
    \end{equation}
    and expand the solution in transverse vector harmonics on $S^{3}$:
    \begin{equation}
    	A^{i}(r,\Omega)=\sum_{n}\!\left(A^{i}_{n}\,r^{n}+\frac{B^{i}_{n}}{r^{\,n+1}}\right)
    	Y^{(\mathrm T)}_{n}(\Omega).
    \end{equation}
    Writing $r=e^{\tau}$, this becomes
    \begin{equation}
    	A^{i}(\tau,\Omega)=\sum_{n=0}^{\infty}
    	\left(A^{i}_{n}\,e^{n\tau}+B^{i}_{n}\,e^{-(n+1)\tau}\right)Y^{(\mathrm T)}_{n}(\Omega).
    	\label{modes} 
    \end{equation}
    The $B^{i}_{n}$ modes are non–normalizable and are set to zero. In the quantum
    theory, $A^{i}_{n}$ create quanta of scaling dimension $n$. We take the equal–$\tau$
    canonical brackets (Dirac brackets in Coulomb gauge) for the transverse modes so that
    \begin{equation}
    	\big[A^{i}_{n},A^{j\dagger}_{n'}\big]=\delta_{nn'}\,\delta^{ij},
    \end{equation}
    which ensures that all physical states have positive norm (only the two transverse
    polarizations propagate).
    
    It is convenient to organize states by the dilatation operator $D$ (the Hamiltonian
    in radial quantization). In our conventions,
    \begin{equation}
    	D=\frac{c\alpha'^{\,2}}{2}\,p^{\mu}p_{\mu}
    	\;+\;\sum_{n>0} n\,A^{i\dagger}_{n}A^{i}_{n},
    \end{equation}
    so that
    \begin{equation}
    	e^{-D t}\,A^{i}_{n}\,e^{D t}=A^{i}_{n}\,e^{-n t}.
    \end{equation}
    Finally, in the abelian $U(1)$ theory the Faddeev–Popov determinant is field–independent
    ($\det(-\nabla^{2})$ in Coulomb gauge); the ghost action is quadratic and decoupled,
    so ghosts can be integrated out and do not affect correlators or the unitarity analysis.
    Thus, thanks to the Coulomb gauge choice—which enforces transversality—the Hilbert space
    contains only positive–norm excitations, establishing unitarity in $D=4$.

%
%
%
%

   Note that  the modes given by Eq.\ref{modes} are centred at $r = 0$, however these modes could be centred anywhere in $4D$. $\tau$ is invariant under the $SO(4)$, hence we can use it to parametrize the motion of the center. The contribution to action is then $\sim \int  d\tau \frac{1}{2c\alpha^2}\dot{x}^\mu (\tau) \dot{x}_\mu (\tau)$. This term is invariant under scale transformation as scale transformation only acts as a translation in $\tau$.  The translation     in $\tau$ by an amount $t$ will be accomplished by $e^{-\frac{c\alpha^2}{2}p^\mu p_\mu t}$. This is  why $\frac{c\alpha^2}{2}p^\mu p^\mu$ appears in $D$. The primary states in the theory are one's obeying
 
 \begin{eqnarray}
 D |a \rangle = a|a \rangle 
 \end{eqnarray}
 Consider a state ${A^\mu}^\dagger_1 {A^\dagger_1}_\mu |0, p^\mu \rangle$. Here, the index $\mu$ is implicitly restricted to the two transverse spatial 
 directions, consistent with the Coulomb gauge choice $A_0=0$ and $\partial_i A^i=0$. 
 Thus only physical oscillators contribute, implying the state is in a representation of $SO(2)$. However the Euclidean symmetry group is $SO(4)$. The only way this is possible if $p^\mu p^\mu = 0$.  This sets $a = -2$. However, ${A^\mu}^\dagger_1   |0, p^\mu \rangle$ by that logic should be  a massless vector, however $p^\mu p^\mu \neq 0$ for this vector. This means, this vector has to not exist in order for the theory to make sense. One can check that the Lagrangian of the theory is invariant under $A^\mu (\bar{x}) \rightarrow -A^\mu (\bar{x})$. One can ask this symmetry is also present at the level of the states, in such a case, the vector ${A^\mu}^\dagger_1   |0, p^\mu \rangle$ will be eliminated from the theory as it is not invariant under $A^\mu (\bar{x}) \rightarrow -A^\mu (\bar{x})$. 
 We also have  that the for a state ${A^\mu}^\dagger_1.. {A^\dagger}\mu_{2n} |0, p^\mu \rangle $ has $\frac{c\alpha^2}{2}p^\mu p^\mu = 2-2n$ implying $M^2 \sim J$ for large $J$ as required. One can check that the states ${A^\mu}^\dagger_1.. {A^\dagger}\mu_{2n} |0, p^\mu \rangle $ are in one to one correspondence with the operators  $  F_{\mu_1 \rho_1} F_{\rho_1 \nu_1}...  F_{\mu_n \rho_n} F_{\rho_n \nu_n} (\vec{x_i}) e^{-i    \int d^4 x   \frac{ \partial^{\nu} F_{\mu\nu}(x) }{|x-x_i|}  }  $   mentioned above.  {All these operators   are constructed solely from $F_{\mu\nu}$, ensuring BRST closure since they commute with the nilpotent BRST charge. In abelian gauge theory the Faddeev--Popov ghosts appear as decoupled quadratic terms in Coulomb gauge and contribute no correlators in the physical sector. Consequently, matrix elements between states created by  these operators are gauge independent, and the physical Hilbert space consists of positive-norm transverse modes. This guarantees unitarity and the absence of negative-norm ghost states.}

 In a CFT, states are in one to one correspondence with corresponding operators at the $r=0$. Hence if we have to evaluate scattering amplitudes of particles, we can simply consider the correlation functions of the corresponding operators. As mentioned above, the fact that the origin $r=0$ is not special,  and the $r=0$ could be moved anywhere in $4D$ tells us we have to integrate over all positions in $4D$ of these operators. However because of the conformal symmetry in the problem, not all these positions are equivalent, hence one has to divide by the volume of the conformal group. This is exactly how we arrived at the Virasoro Shapiro amplitude above.

 \section*{Discussion}
 
 In this work we have shown that the characteristic analytic structure of the
 Virasoro--Shapiro amplitude can be reproduced within a purely four-dimensional
 conformal framework, using only general field-theoretic ingredients such as
 conformal invariance, the operator--state correspondence in radial quantization,
 and analyticity of correlation functions. The construction is entirely
 kinematical: it does not rely on worldsheet geometry, string degrees of freedom,
 or a spacetime interaction Lagrangian.
 
 Several implications follow.
 
 \begin{itemize}
 	\item
 	The construction provides an explicit four-dimensional realization of the
 	analytic structure --- crossing symmetry, Regge behavior, and an infinite tower
 	of exchanged states --- that is usually associated with string amplitudes. The
 	appearance of this structure should be understood as a consequence of symmetry
 	and consistency rather than of spacetime interactions in a local quantum field
 	theory.
 	
 	\item
 	The result demonstrates that string-like analytic behavior need not be tied to a
 	two-dimensional worldsheet description or to critical string dynamics. Instead,
 	it can arise from conformal kinematics in four dimensions, suggesting the
 	possibility of string-like scattering structures in non-string ultraviolet
 	completions.
 	
 	\item
 	From the perspective of the S-matrix bootstrap, the construction supplies a
 	concrete, self-consistent target amplitude whose analytic structure saturates
 	crossing and Regge constraints in four dimensions. As such, it provides a useful
 	benchmark for bootstrap approaches that seek to characterize consistent
 	high-energy scattering purely from general principles.
 \end{itemize}
 
 
 Open questions remain, most notably the identification of a microscopic
 completion whose conformal data reproduces the correlators used here. If such a
 completion exists, it is expected to be strongly coupled  and its
 explicit construction lies beyond the scope of the present work. Nonetheless,
 the results presented here indicate that the Virasoro--Shapiro analytic structure
 can emerge naturally in four dimensions without invoking strings, and may serve
 as a guide in the search for new conformal fixed points or kinematical frameworks
 exhibiting string-like scattering behavior.

\end{document}